\documentclass[useAMS,usenatbib]{mn2e}
\usepackage{psfrag,graphicx}
\usepackage{color}
\usepackage{txfonts}
\usepackage{breqn}
\usepackage[T1]{fontenc}
\usepackage{ae,aecompl}

\title[Black hole formation]
{Witnessing the birth of a supermassive protostar}
 \author[Latif  et al.]
{M. A. Latif\thanks{Corresponding author: latif@iap.fr}$^{1,2}$,
D. R. G. Schleicher$^{3}$,
T. Hartwig$^{1,2}$\\
$^1$ Sorbonne Universités, UPMC Univ Paris 06, UMR 7095, Institut d'Astrophysique de Paris, F-75014, Paris, France \\
$^2$CNRS, UMR 7095, Institut d'Astrophysique de Paris, F-75014, Paris, France \\
$^3$ Departamento de Astronomía, Facultad Ciencias Físicas y Matemáticas, Universidad de Concepción, \\
       Av. Esteban Iturra s/n Barrio Universitario, Casilla 160-C, Chile \\
}
\date{}

\def\LaTeX{L\kern-.36em\raise.3ex\hbox{a}\kern-.15em
      T\kern-.1667em\lower.7ex\hbox{E}\kern-.125emX}

\begin{document}

\bibliographystyle{mn2e}

\label{firstpage}

\maketitle
\def\na{NewA}%
          % New~Astronomy
\def\aj{AJ}%
          % Astronomical Journal
\def\araa{ARA\&A}%
          % Annual Review of Astron and Astrophys
\def\apj{ApJ}%
          % Astrophysical Journal
\def\apjl{ApJ}%
          % Astrophysical Journal, Letters
\def\jcap{JCAP}

\def\apjs{ApJS}%
          % Astrophysical Journal, Supplement
\def\ao{Appl.~Opt.}%
          % Applied Optics
\def\apss{Ap\&SS}%
          % Astrophysics and Space Science
\def\aap{A\&A}%
          % Astronomy and Astrophysics
\def\aapr{A\&A~Rev.}%
          % Astronomy and Astrophysics Reviews
\def\aaps{A\&AS}%
          % Astronomy and Astrophysics, Supplement
\def\azh{AZh}%
          % Astronomicheskii Zhurnal
\def\baas{BAAS}%
          % Bulletin of the AAS
\def\jrasc{JRASC}%
          % Journal of the RAS of Canada
\def\memras{MmRAS}%
          % Memoirs of the RAS
\def\mnras{MNRAS}%
          % Monthly Notices of the RAS
\def\pra{Phys.~Rev.~A}%
          % Physical Review A: General Physics
\def\prb{Phys.~Rev.~B}%
          % Physical Review B: Solid State
\def\prc{Phys.~Rev.~C}%
          % Physical Review C
\def\prd{Phys.~Rev.~D}%
          % Physical Review D
\def\pre{Phys.~Rev.~E}%
          % Physical Review E
\def\prl{Phys.~Rev.~Lett.}%
          % Physical Review Letters
\def\pasp{PASP}%
          % Publications of the ASP
\def\pasj{PASJ}%
          % Publications of the ASJ
\def\qjras{QJRAS}%
          % Quarterly Journal of the RAS
\def\skytel{S\&T}%
          % Sky and Telescope
\def\solphys{Sol.~Phys.}%
          % Solar Physics

          % Solar Physics
\def\sovast{Soviet~Ast.}%
          % Soviet Astronomy
\def\ssr{Space~Sci.~Rev.}%
          % Space Science Reviews
\def\zap{ZAp}%
          % Zeitschrift fuer Astrophysik
\def\nat{Nature}%
          % Nature
\def\iaucirc{IAU~Circ.}%
          % IAU Cirulars
\def\aplett{Astrophys.~Lett.}%
          % Astrophysics Letters
\def\apspr{Astrophys.~Space~Phys.~Res.}%
          % Astrophysics Space Physics Research
\def\bain{Bull.~Astron.~Inst.~Netherlands}%
          % Bulletin Astronomical Institute of the Netherlands
\def\fcp{Fund.~Cosmic~Phys.}%
          % Fundamental Cosmic Physics
\def\gca{Geochim.~Cosmochim.~Acta}%
          % Geochimica Cosmochimica Acta
\def\grl{Geophys.~Res.~Lett.}%
          % Geophysics Research Letters
\def\jcp{J.~Chem.~Phys.}%
          % Journal of Chemical Physics
\def\jgr{J.~Geophys.~Res.}%
          % Journal of Geophysics Research
\def\jqsrt{J.~Quant.~Spec.~Radiat.~Transf.}%
          % Journal of Quantitiative Spectroscopy and Radiative Trasfer
\def\memsai{Mem.~Soc.~Astron.~Italiana}%
          % Mem. Societa Astronomica Italiana
\def\nphysa{Nucl.~Phys.~A}%
          % Nuclear Physics A
\def\physrep{Phys.~Rep.}%
          % Physics Reports
\def\physscr{Phys.~Scr}%
          % Physica Scripta
\def\planss{Planet.~Space~Sci.}%
          % Planetary Space Science
\def\procspie{Proc.~SPIE}%
          % Proceedings of the SPIE

% 

% \newcommand{\ch}[1]{\textcolor{red}{\textbf{#1}}}
% \newcommand{\newln}{\\&\quad\quad{}}
% \date{today}

 \begin{abstract}
 {The detection of $\rm z>6$ quasars reveals the existence of supermassive black holes of  a few $\rm 10^9~M_{\odot}$. One of the potential pathways to explain their formation in the infant universe is the so-called  direct collapse model which provides massive seeds of $\rm 10^5-10^6~M_{\odot}$. An isothermal direct collapse mandates that halos should be of a primordial composition and the formation of molecular hydrogen remains suppressed in the presence of a strong Lyman Werner flux.
 In this study, we perform high resolution cosmological simulations for two massive primordial halos employing a detailed chemical model which includes  $\rm H^-$ cooling as well as realistic opacities for both the bound-free  $\rm H^-$ emission and the Rayleigh scattering of hydrogen atoms. We are able to resolve the collapse up to unprecedentedly high densities of  $\rm \sim 10^{-3}~g/cm^3$ and to scales of about $\rm 10^{-4}$ AU.  Our results show that the gas cools down to  $\rm \sim $ 5000 K in the presence of $\rm H^-$ cooling, and induces fragmentation  at scales of about 8000 AU in one of the two simulated halos, which may lead to the formation of  a binary. In addition, fragmentation also occurs on the AU scale in one of the halos but the clumps are expected to merge on short time scales. Our results confirm that $\rm H^-$ cooling  does not prevent the formation of a supermassive star and the trapping of cooling radiation stabilises the collapse on small scales.} 
 \end{abstract}
% context
% Hitherto, cosmological simulations conducted to study the direct collapse  ignored $\rm H^-$ cooling, expected to occur at densities $\rm > 10^8 ~cm^{-3}$ and  also arbitrarily switched off cooling  at a given density. In this study, we perform high resolution cosmological simulations for two massive primordial halos by employing a detailed chemical model which includes $\rm H^-$ cooling as well as realistic opacities for  both the bound-free $\rm H^-$ emission and the Rayleigh scattering of hydrogen atoms.
%  results

% conclusion 

\section{Introduction} \label{sec:intro}
Supermassive black holes of a few million to billion solar masses not only reside in the centres of present day galaxies, but there is compelling evidence for their existence  at z>6 from the observations of  high redshift quasars  \citep{Fan2006,Willott2007,MOrtlock2011,Venemans2013,Banados2014,Venemans2015,Wu2015}. Through the stacking of X-ray observations, the accreted black hole mass density at z $\rm \sim$ 6 has been estimated to be $\rm  \lesssim 1000 ~M_{\odot}~Mpc^{-3}$  \citep{Treister2013}. The formation of such massive objects within the first billion years after the Big Bang is  still a mystery.  Various methods have been proposed in the literature  to explain their existence such as  stellar mass  black holes from population III stars \citep{Johnson2007,Alvarez2009,LatifPopIII13,Madau2014},  seed black holes forming from the dynamical processes in dense stellar clusters \citep{Baumgarte1999, Devecchi2009,Lupi2014} and the collapse of protogalactic gas clouds into a  massive object, the so-called direct collapse model \citep{Rees78,Spaans2006,Volonteri2008,Latif2011,Latif2013c,Ferrara14}. Dedicated reviews on these topics are given by  \cite{Volonteri2010} and \cite{Haiman2013}.

 The direct collapse scenario has been a topic of great interest during the past decade as it provides massive seeds of about $\rm 10^{5}-10^6~M_{\odot}$ which are two orders of magnitude more massive than black hole seeds resulting from other scenarios. The prime requirement for the direct collapse is the presence of large inflow rates of $\rm \geq 0.1 ~M_{\odot}/yr$ to assemble a supermassive  star which later collapses into a direct collapse black hole (DCBH) via relativistic instabilities \citep{Begelman2010,Schleicher13,Hosokawa2013,Sakurai2015,Pacucci2015}.  Such large inflow rates can be obtained either via dynamical processes such as the bars-within-bars instability \citep{Begelman2006,Begelman2009} or thermodynamically by keeping the gas warm which results in large sound speed  and consequently higher inflow rates \citep{Schleicher10,Latif2013d,Johnson2013b,Latif2014UV,Yue2014}.  The latter mandates that the host halo should be metal free and illuminated by a strong Lyman Werner (LW) flux for which molecular cooling and consequently, fragmentation remains suppressed \citep{Omukai2001,Shang2010,Regan2014B,Visbal2014,Latif2015a}.  The complete dissociation of  molecular hydrogen requires the presence of a  strong  LW flux above a certain critical strength which depends on the spectra of the stars \citep{Sugimura14,Dijksta2014,Agarwal2015,Latif2015a,Glover2015b,Hartwig2015}.
 
 Numerous numerical experiments have been performed to study the direct collapse scenario and show that in the presence of an intense LW flux the formation of molecular hydrogen remains suppressed and the collapse occurs isothermally with T $\sim$ 8000 K \citep{Brom2003,Wise2008,Regan09,Latif2013c,Prieto2013,LatifMag2014,Regan2014a,Choi2015,Bcerra2014}. Moreover, these simulations demonstrate that fragmentation occurs occasionally but large inflow rates of about $\rm 0.1-1~M_{\odot}/yr$ are available to grow massive objects in these conditions. \cite{Latif2013d} and \cite{Shlosman2015} employed sink particles to follow the long term evolution finding that massive object of about $\rm 10^5~M_{\odot}$ can be formed within a million years which confirms the feasibility of  the direct collapse model. The role of magnetic fields during the direct collapse scenario was explored  by \cite{Latifdynamo2013} and \cite{LatifMag2014}  finding  that  seed fields get amplified during the collapse via turbulent dynamo and reach the equipartition field strength. Such strong magnetic fields help in suppressing fragmentation and transferring the angular momentum. 
 
Starting from cosmological initial conditions, these simulations were able to resolve the collapse  down to AU scales but  did not employ a detailed chemical model. These studies  ignored $\rm H^-$ cooling  which is effective at densities between $\rm 10^{-16}-10^{-8} ~g/cm^{3}$, and it can cool the gas down to a few thousand K, and  hence may induce fragmentation \citep{Omukai2001,VanBorm2014}.  To mimick the formation of a protostar, cooling has been switched off in previous simulations above a critical density \citep{Latif2013c,Regan2014a,Bcerra2014}, leading to an adiabatic evolution. They found that although fragmentation is not completely suppressed, large inflow rates may lead to the formation of a supermassive star. On the other hand, improving on previous works \cite{Inayoshi2014} and \cite{VanBorm2014} performed 3D simulations starting from idealised initial conditions (a spherical gas cloud with a top hat  density profile) which included turbulence and rotation as well as  $\rm H^-$ cooling and  opacities for  the $\rm H^-$ bound-free emission, free-free absorption and the Rayleigh scattering of hydrogen atoms. They found that no strong fragmentation occurs under these conditions. However,  it is not yet clear what will be the impact of these processes in cosmological simulations compared to the adhoc initial conditions.

In this study, we perform the first cosmological simulations resolving the collapse to unprecedentedly high densities  by employing  a detailed chemical model which includes all  relevant chemical processes. Particularly, our simulations include $\rm H^-$ cooling as well as  realistic opacities for  both atomic  and molecular hydrogen cooling which become relevant at densities $\rm > 10^{-16}~g/cm^{3}$.  To achieve this goal, we took two  massive primordial halos of a few times $\rm 10^7~M_{\odot}$  and followed their collapse down to about $\rm 10^{-4}$ AU and $\rm 2 \times 10^{-3}$ AU, respectively.  To resolve turbulent eddies, we employed a fixed Jeans resolution of 64 cells during the entire course of simulations.  Our results show that  the formation of massive objects is plausible in the presence of $\rm H^-$ cooling and realistic opacities help stabilising the collapse on small scales.
  
This article is structured as follows. In section 2, we discuss the  initial conditions, provide the details of the numerical methods and the chemical model employed in this work.  We present our main results in section 3. In section 4, we summarise our main conclusions.

\section{Numerical methods}
The simulations presented here are conducted using the open source code  Enzo\footnote{http://enzo-project.org/} \citep{OShea2004,Enzocode2014} which is an adaptive mesh refinement, parallel,  Eulerian grid based cosmological simulations code. It can be ported to various platforms and has been widely used for high resolution simulations.  We employed the Runge Kutta based MUSCL scheme to solve the hydrodynamical equations which is  a second-order modified form of the Godunov's method. The multi-grid Poisson solver is used for computing self-gravity. The dark matter (DM) dynamics is solved using the particle-mesh technique. 
 
\subsection{Simulations setup}
We start our simulations from cosmological initial conditions at z=100 and a top grid resolution of $\rm128^3$ cells ($\rm128^3$ DM particles). Our computational box has a comoving size of 1~Mpc/h. We first run DM only simulations to select the most massive halo in our simulation volume at z=15. We rerun the simulation  with the most massive halos at the center of the box and employ two additional nested refinement levels each with a grid resolution of  $\rm128^3$ cells and $\rm128^3$ DM particles.  In total, we use 5767168 DM particles which provide an effective DM resolution of about 600 $\rm M_{\odot}$. To avoid numerical artefacts, we smooth DM particles at the 12th refinement level i.e., 2.7 pc in comoving units.  We in total add 40 (36 for the halo 2) dynamic refinement levels during the simulations in the central 62 kpc (comoving units) which gives us an effective spatial resolution of $\rm 10^{-4}$ AU ($\rm 2 \times10^{-3}$ AU for the halo 2). Our refinement criterion is based on the baryon density, the DM mass resolution and a fixed Jeans resolution of 64 cells during the entire course of the simulations. We refine the grid cells exceeding four  times the  cosmic mean density or the DM particle density  0.0625 times  $\rho_{DM}r^{\ell \alpha}$  where $\rho_{DM}$ is the dark matter density, r = 2 is the refinement factor, $\ell$ is the refinement level, and $\alpha = -0.3$ makes the refinement super-Lagrangian. We use the WMAP 7 year data to generate cosmological initial conditions \citep{Jarosik2011}.

We select two halos of $\rm 4 \times 10^7~M_{\odot}$ (halo 1) and $\rm 6 \times 10^7~M_{\odot}$ (halo 2) and assume that they are metal free.  The properties of the halos are listed in table \ref{table1}. The collapse redshifts of halo 1 and halo 2  are 10.7 and 10.1, respectively. Halo 2 is more massive, collapses about 10 Myrs later compared to  halo 1 and has a higher viral temperature.  Halo 2 has a  slightly higher spin due to the later collapse time.

\cite{Latif2015dust} found that about 50 \% of halos within this mass range are metal free at z=10.   We assume that  these halos are illuminated by a uniform strong LW flux of strength $\rm 10^5$ in units of $\rm J_{21}$ where $\rm J_{21}=1$ is $\rm 10^{-21}~ erg/cm^2/s/Hz/sr$. One-zone calculations of \cite{Sugimura14} and \cite{Agarwal2015} suggest that the value of $\rm J_{crit}$ for  realistic spectra of the first galaxies is about 1000. However, 3D cosmological simulations show that one-zone models often underestimate the value of $\rm J_{crit}$ due to their inability to model the collapse dynamics and hydrodynamical  effects including in particular the virial shock \citep{Shang2010,Latif2014UV,Regan2014B,Latif2015a}. The strength of the LW flux ($\rm J_{21}=10^5$) employed here is about a factor of two above the $\rm J_{crit}$ found from 3D cosmological simulations by \cite{Latif2014UV,Latif2015a} for realistic spectra of the first galaxies. This choice of the flux ensures that the collapse occurs isothermally and molecular hydrogen cooling remains suppressed.
The LW flux with $\rm T_{rad}= 2 \times 10^4$ K  is turned on at  z=30. We stop our simulations when a peak density of $\rm 2 \times10^{-3}~g/cm^{3}$ and $\rm 10^{-5}~g/cm^{3}$ is reached in halo 1 and halo 2, respectively.  At these densities the protostar begins to form and the collapse is expected to evolve adiabatically during the later stages.

 \begin{table*}.
\begin{center}
\caption{Halo properties.}
\begin{tabular}{ccccccccc}
\hline
\hline

Model   & Mass   & Redshift  & $R_{\rm vir}$  & spin parameter &Gravitational energy &$V_{rms}$ &$T_{vir}$\\

No & $\rm M_{\odot}^{DM}, \rm M_{\odot}^{gas}$ & $z$  &  kpc  & $ \lambda_{DM}, \lambda_{gas}$  & erg/s & km/s & K\\
\hline                                                          \\

1     & $\rm 4 \times 10^{7}, \rm 7.7 \times 10^{6}$ &10.7  &1.3  &0.03, 0.028&$\rm 6.37 \times 10^{52}$ & 18 & $\rm 1.3 \times 10^{4}$\\
2     & $\rm 6 \times 10^{7}, \rm 1.1 \times 10^{7}$ &10.1  &1.49  &0.034, 0.033&$\rm1.25 \times 10^{53}$ &21 & $\rm 1.8 \times 10^{4}$\\
%D     & $\rm 1.42 \times 10^{7}$  & 14.23 &$T_{\rm rad}= 10^4$      & $J_{21}=$600    &0.025 \\
%E     & $\rm 2.4 \times 10^{7}$   & 12.95 &$T_{\rm rad}= 10^4$     & $J_{21}=$1500    &0.009 \\
\hline
\end{tabular}
\label{table1}
\end{center}
\end{table*}

\subsection{Chemical model}
We employ the chemical package KROME  \citep{Grassi2014} to solve the thermal and chemical evolution of the gas along with cosmological simulations.  We solve the rate equations of the following primordial species $\rm H,~H^+,~ H^-, ~He,~ He^+, ~He^{++},~ H_2, ~H_2^+, ~e^-$  and couple them with hydrodynamics. We ignore the deuterium  species as they can only can be relevant  when gas temperature is about 200 K. Furthermore, the HD molecule easily gets dissociated even in the presence of a weak LW flux.  The chemical network used in this study is listed in the appendix table 1 of \cite{Latif2015a}.  In our model, we include the formation, the photo-dissociation  and the collisional dissociation of molecular hydrogen,  the photo-detachment of H$^-$, the photo-dissociation of $\rm H_{2}^+$ as well as collisional ionisation and radiative recombination reactions of various species. We implemented a comprehensive model of cooling/heating which consists of cooling due to the collisional excitation, collisional ionization, radiative recombination, Bremsstrahlung radiation, molecular hydrogen, collision induced emission, chemical heating and cooling from three-body reactions. For molecular hydrogen cooling, we use the expression for the escape probability provided by \cite{Omukai2000} and employ the $\rm H_2$ self-shielding fitting formula given by \cite{WolocottGreen2011}.

For this study, as we explore the high density regime, we also include $\rm H^-$ cooling  which is due to the energy liberated in the form of a photon produced during the radiative association of hydrogen atoms with electron. The net cooling rate can be approximated as \citep{Omukai2001,Schleicher10,VanBorm2014}:
\begin{equation}
  \Lambda_{H^-} \approx k_{H^-} n_{H}n_e E_{\gamma}
\label{eq0}
\end{equation} 
where E$_{\gamma}$ is the energy carried by a photon, which is approximately $\rm E_{0} + k_B T$ where $\rm E_0$ is the binding energy of H$^-$ $\sim 0.75~eV$ and  $\rm k_{B}$ is the Boltzmann constant.  At densities of about $\rm 10^{-8}~g/cm^3$,  the gas cloud becomes optically thick  both to the  H$^-$ bound-free emission and the Rayleigh scattering of hydrogen atoms. Consequently, cooling is suppressed as both of these are important coolants in the high density regime. To take into account these effects, we apply optical depth approximations both for the Rayleigh scattering of H atoms and the H$^-$ bound free absorption.  The optical depth correction for the Rayleigh scattering of H atoms is given as \citep{Omukai2001,VanBorm2014}
\begin{equation}
 \tau_{RS} = \sigma_{RS}n_{H} \lambda_J/2
\label{eq1}
\end{equation} 
where $\rm n_H$ is the number density of H atoms, $\lambda_J$ is the Jeans length and  $\sigma_{RS}$ is the Rayleigh scattering cross-section  for hydrogen given as 
\begin{equation}
 \sigma_{RS} =  5.799 \times 10^{-29} \lambda^{-4} + 1.422  \times 10^{-30} \lambda^{-6} + 2.784 \times 10^{-22} \lambda^{-8} ~cm^2
\label{eq2}
\end{equation} 
where $\lambda$ is the wavelength in $\mu m$. The collisional excitation cooling rate is multiplied by $ {\rm exp}(-\tau_{RS})$. Further details about these processes are given in \cite{Omukai2001} and \cite{VanBorm2014}. At densities between $\rm 10^{-7}- 10^{-5}~g/cm^{3}$, collisional ionisation  cooling becomes operational and keeps the gas temperature of about $\rm 10^4$ K. Above $\rm 10^{-4}~g/cm^{3}$, the collapse proceeds adiabatically and the protostar is expected to form at this stage.

\section{Results}
We present here the main findings of this study. In the following subsections, we discuss the thermal, chemical and dynamical properties of the halos at the final state of our simulations.  In the end, we estimate the amount of fragmentation and discuss  the impact of $\rm H^-$ cooling.

\subsection {Thermal and chemical properties}
\begin{figure*}
\includegraphics[scale=0.4]{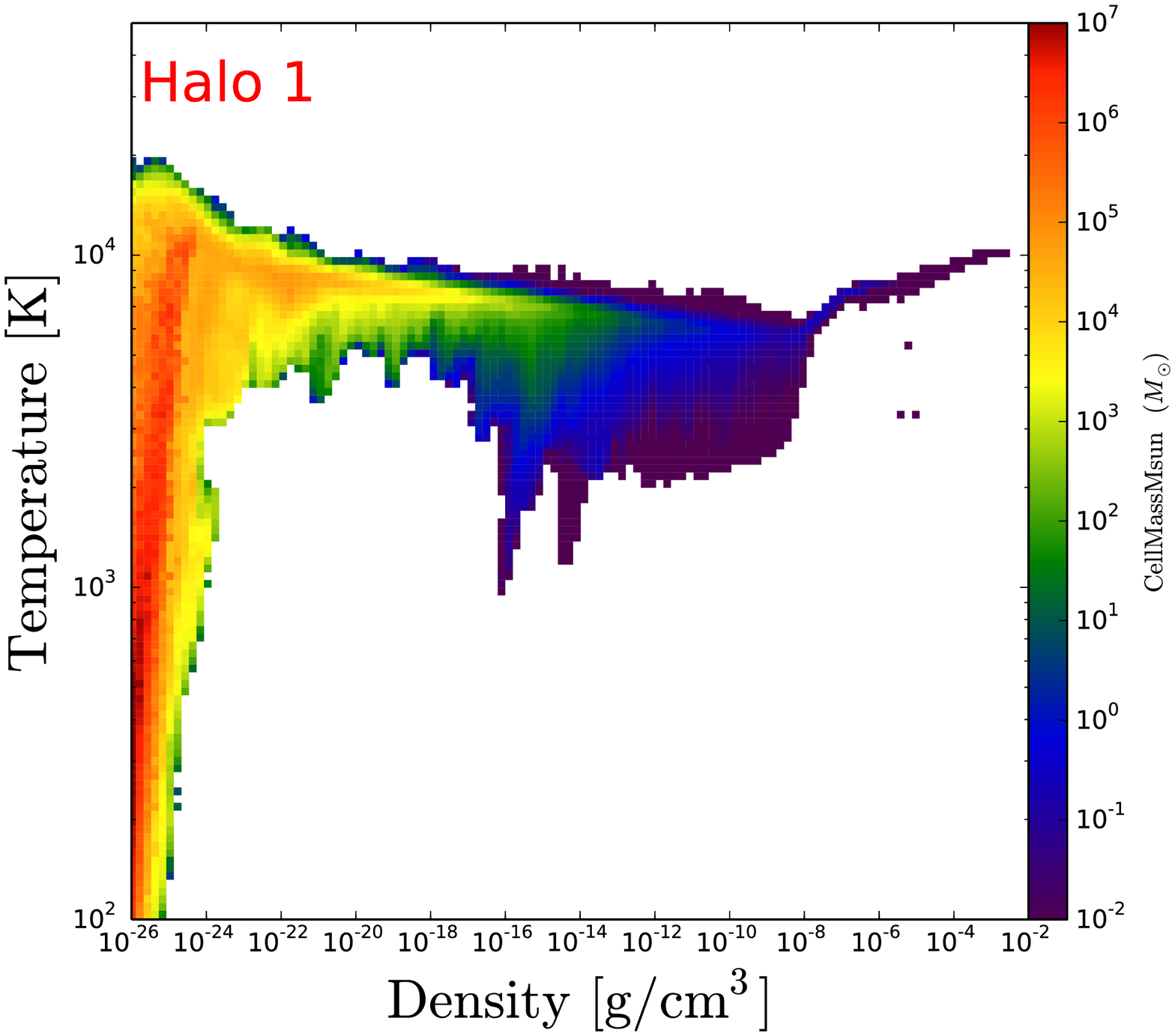}
\includegraphics[scale=0.4]{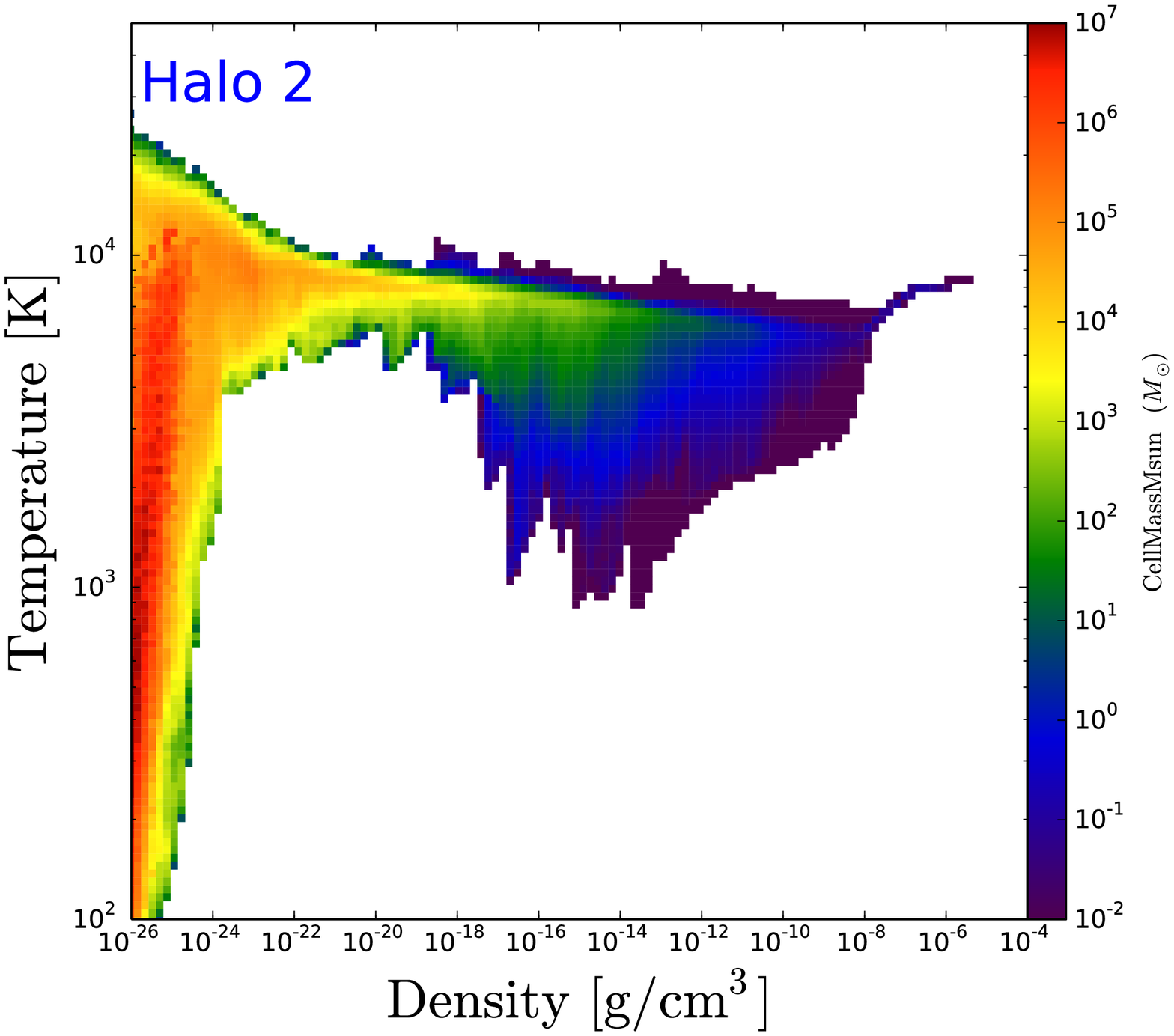} 
\includegraphics[scale=0.7]{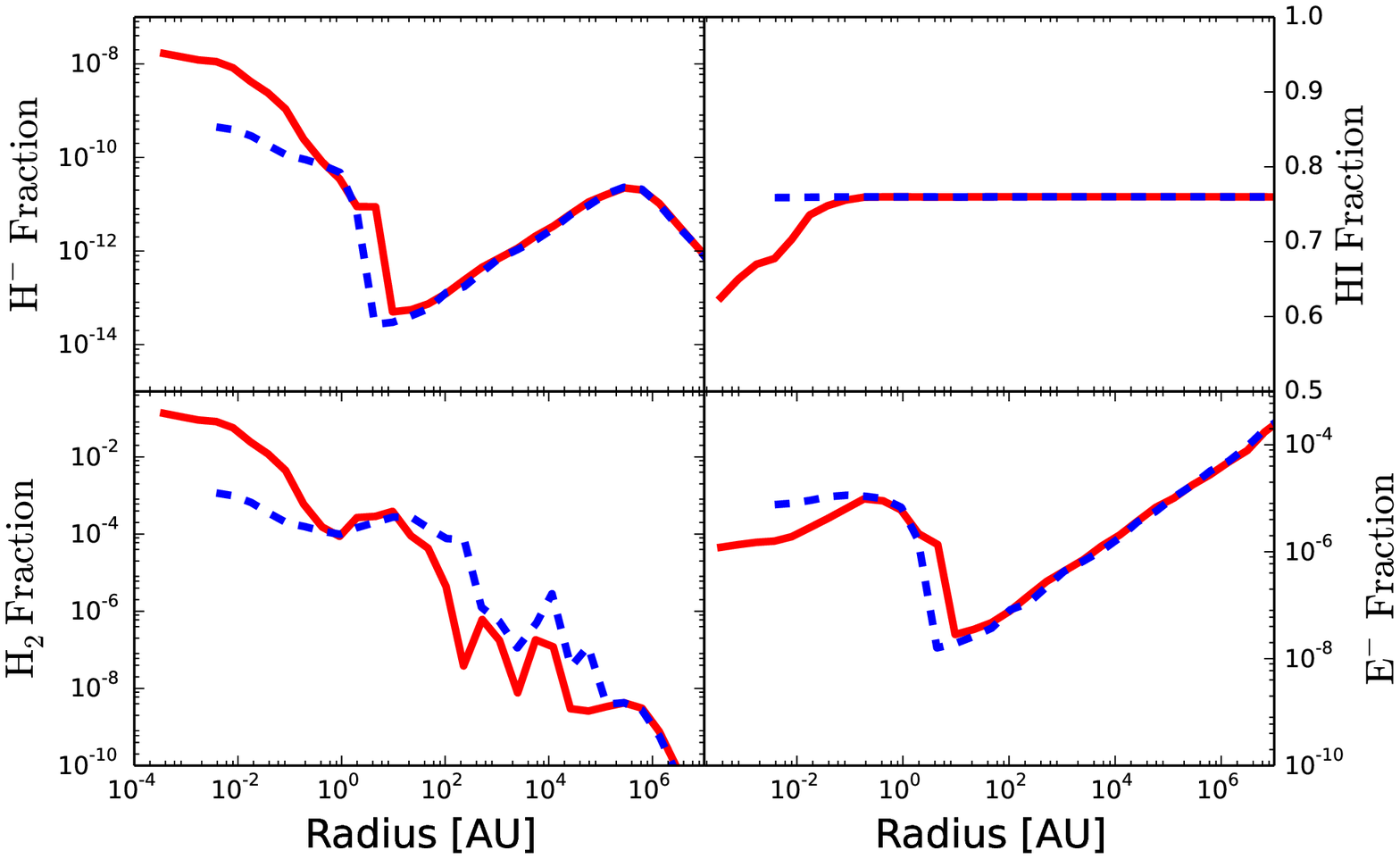}
\caption{The density-temperature phase diagram  and the spherically averaged abundances of chemical species for halo 1 (red line) and halo 2 (blue line) at the end of our simulations. The figure shows that the gas in the halos collapses almost isothermally up to densities of $\rm 10^{-16}~g/cm^3$ and H$^-$ cooling becomes effective at densities $\rm >10^{-16}~g/cm^3$ and cools the gas down to about 5000 K.  Above densities of $\rm 10^{-8}~g/cm^3$, the gas becomes optically thick to both the $\rm H^-$ and the collisional excitation cooling and the temperature starts to increase. The collisional ionisation cooling keeps the gas temperature to $\rm 10^4$ K up to $\rm10^{-4}~g/cm^3$ and thereafter the collapse is expected to proceed adiabatically as all cooling channels shut down.} 
\label{fig0}
\end{figure*}

 The halos in our simulations are assembled via the gravitational collapse of density perturbations in  the DM potentials. They further grow via accretion from the cosmic web and the merging of minihalos of about $\rm10^5-10^6 ~M_{\odot}$. In the presence of an intense LW flux,  cooling due to the molecular hydrogen remains suppressed and the halo mass reaches  the atomic cooling threshold. The masses  of the halo 1 and halo 2 are  $\rm 4 \times 10^7~M_{\odot}$ and $\rm 6 \times 10^7~M_{\odot}$ and the collapse redshifts are 10.7, 10.1, respectively.  As shown in figure \ref{fig0}, the collapse occurs isothermally with a temperature  of about 8000 K under the intense LW flux and the formation of molecular hydrogen remains suppressed. At densities between $\rm 10^{-16}-10^{-8} ~g/cm^{3}$, the $\rm H^-$ cooling  becomes a strong contributor  and brings the gas temperature down to  $\rm \sim$ 5000 K.  At densities of $\rm 10^{-16}~g/cm^3$, the gas becomes optically thick both to the $\rm H^-$ and the collisional excitation cooling of hydrogen atoms, and the gas temperature starts to increase. The collisional ionisation cooling takes over and keeps the temperature close to $\rm 10^4$ K. The  gas becomes completely optically thick to cooling radiation at densities $\rm \geq 10^{-4}~g/cm^3$ and  thereafter the collapse  proceeds adiabatically. The thermal evolution is similar in both halos but the impact of $\rm H^-$ cooling is more prominent in the halo 2 compared to the halo 1. \cite{Inayoshi2014} performed idealised simulations for a collapsing cloud without any LW flux finding that  the gas becomes optically thick to cooling radiation already at $\rm \geq 10^{-8}~g/cm^3$ and the temperature rises to about $\rm 2 \times 10^4$ K. In our case  collisional ionisation cooling proceeds up to $\rm 10^{-4}~g/cm^3$ and keeps the temperature close to $\rm 10^4 K$. In comparison with  \cite{Inayoshi2014},  the temperature is about a factor of two lower at densities of $\rm 10^{-7}~g/cm^3$ as collisional ionisation cooling is not explicitly included in their model. A direct comparison of the thermal properties with  \cite{Regan2014a} and \cite{Bcerra2014} is not possible as they ignored both $\rm H^-$ cooling and realistic opacities for atomic line cooling. At densities $\rm > 10^{-4}~g/cm^3$, we reach the resolution limit of our simulations but we expect a close to adiabatic evolution in this regime. In spite of these differences, our results for the thermal and chemical properties are quite similar to \cite{Inayoshi2014} and \cite{VanBorm2014} because of the self-similar behavior of the isothermal run-away collapse.

The mass fractions of  various chemical species are shown in the bottom panel of figure \ref{fig0}. The gas remains almost neutral with a  mass fraction of 0.75 for both halos while for the halo 1 it declines in the very centre due to the formation of molecular hydrogen. The fraction of molecular hydrogen gets boosted during the collapse at densities $\rm \geq 10^{-6}~ g/cm^3$ due to  the three-body processes.  However, the cooling from molecular hydrogen remains negligible as the gas at these densities is optical thick to  $\rm H_2$ cooling. Both electron and H$^-$ fractions  decrease with radius down to 10 AU due to the decline in the gas temperature and increase again in the centre of the halo as the temperature rises because of the trapping of the cooling radiation.
% \ch{ The molecular hydrogen in \cite{Inayoshi2014} and \cite{Bcerra2014} gets dissociated at densities $\rm \geq 10^{-6}~ g/cm^3$  because of the higher temperatures of a few times $\rm 10^4 $ K. We expect this to happen during the later phases of collapse at densities above $\rm 10^{-4}~g/cm^3$ when collapse evolves adiabatically. Given these differences, our results for the chemical properties of the gas are overall in agreement with their findings.}

\subsection{Physical and dynamical properties}
\begin{figure*}
\includegraphics[scale=0.7]{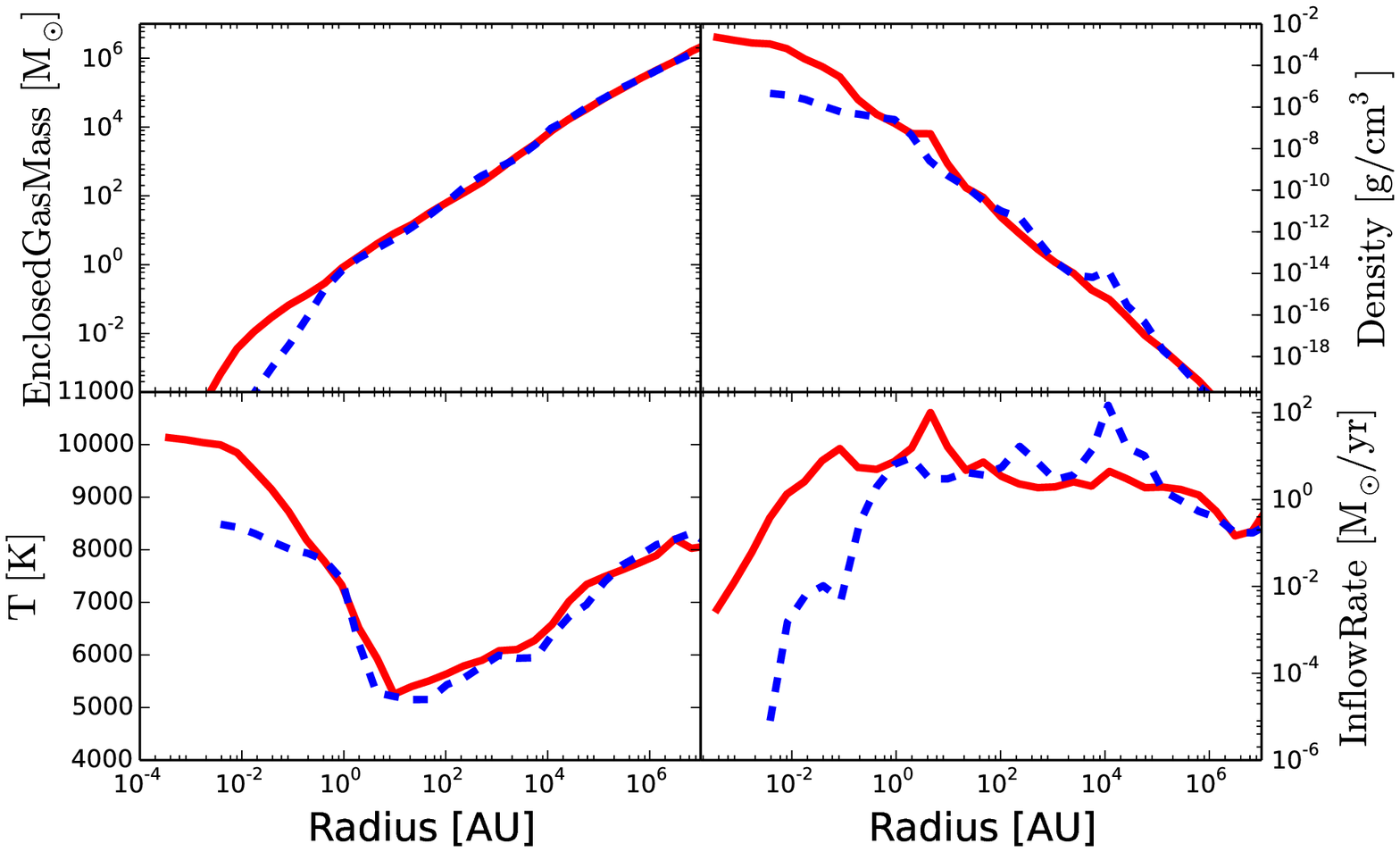}  \\
\hspace{0.8 cm}\includegraphics[scale=0.7]{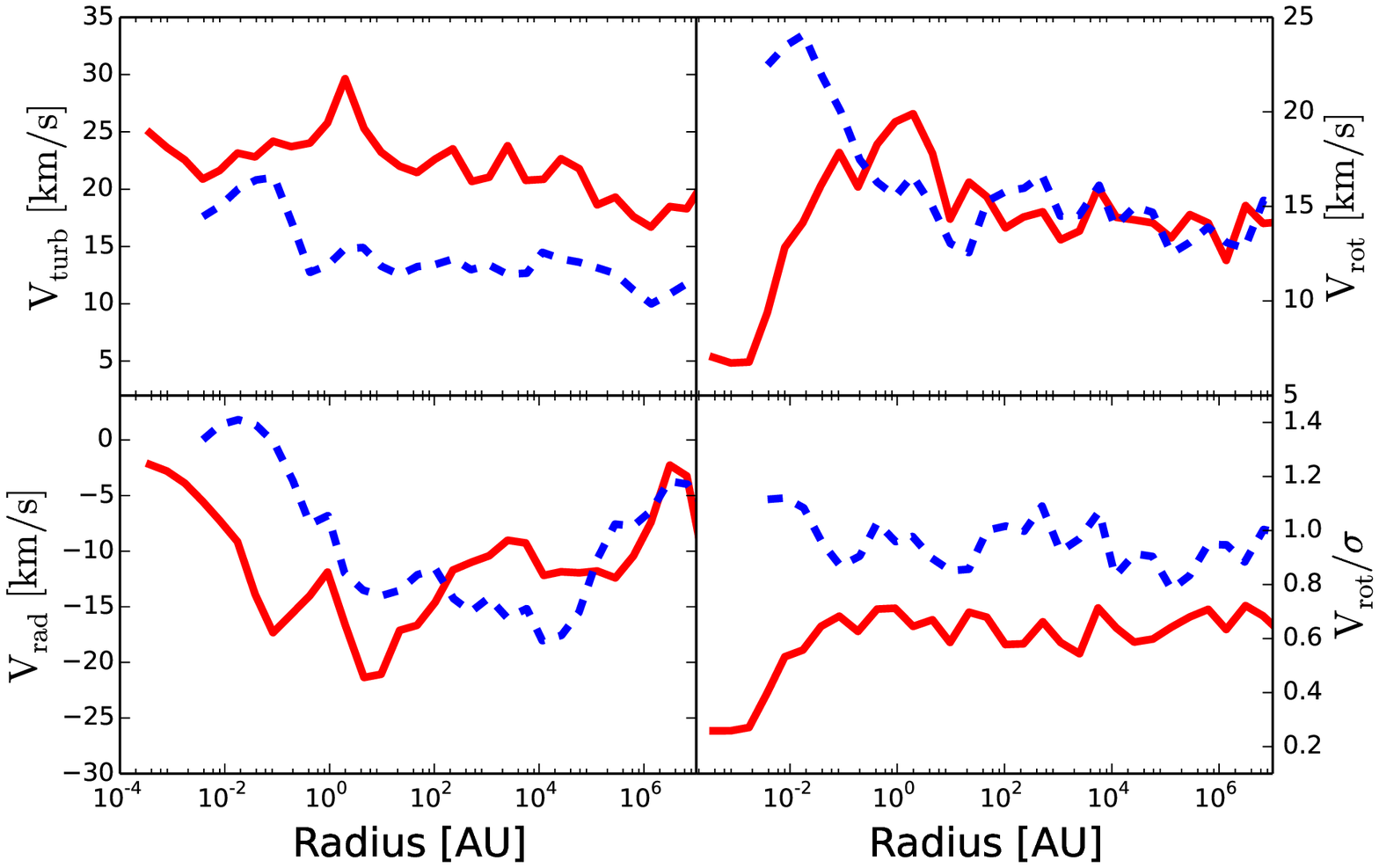} 
\caption{Spherically averaged and radially binned radial profiles at the end stage of our simulations are shown here. In the top panel, the density, the temperature, the enclosed gas mass and the mass inflow rates while in the bottom panel, the radial velocity, the turbulent velocity, the sound speed and  the ratio of V$_{\rm rot}/ \sigma$. The red and blue lines represent halo1 and halo 2, respectively. The figure shows the average behaviour of the above mentioned quantities at various stages of the collapse including the initial isothermal collapse, the H$^-$ cooling and the radiation trapping regimes.}
\label{fig}
\end{figure*}

The peak density reached in our simulations for halo 1 is  $\rm 2 \times 10^{-3}~g/cm^3$ while it is  $\rm \sim 10^{-5}~g/cm^3$ for halo 2 as shown in the top panel of  figure \ref{fig}. The density increases with $\rm R^{-2}$ for both halos as expected for an isothermal collapse \citep{Larson1969,Penston1969} and flattens out within the central core. This is known as the core-envelope structure, where the core has a constant density and the envelope has $\rm R^{-2}$ profile. The bumps in the density profile around 10 AU for halo 1 and  10$^4$ AU for  halo 2 are due to the presence of additional clumps on these scales. The total gas mass within the central 30 pc of the halo is about  $\rm 2 \times 10^6 ~M_{\odot}$  and the mass profile increases linearly with the radius while it drops sharply inside the central core. The average temperature above $\rm \geq 10^5$ AU is about 8000 K, drops down to $\rm \sim $5000 K between 2-100 AU due to the $\rm H^-$ cooling and below 1 AU temperature starts to increase again to 10$^4$ K as the cooling radiations is trapped.

The estimates of the mass inflow rates suggest that large inflows of about $\rm 1~M_{\odot}/yr$ are available down to $\rm 10^{-2}$ AU. The small wiggles on the mass inflow rates are due to the presence of additional clumps in both halos which are  also reflected in the density profiles.  In contrast to the envelope, the mass inflow rate drops in the core for both halos. This comes from the fact that the radial infall velocity depicted in the bottom panel of figure \ref{fig} becomes zero in the core and is high inside the envelope with $\rm \sim$ 15 km/s. This is the characteristic of a runaway collapse and has been observed previously. The typical turbulent velocity is about 10-25 km/s and increases towards the centre due to the larger infall of the gas in the halo centre.  It is about a factor of 1.5 higher in the halo 1 compared to the halo 2. The typical rotational velocity in the envelop is about 15 km/s in both halos and shows a self-similar  behaviour in the envelope as shown by \cite{Saigo1998}. However, in the central core the rotational velocity is higher in halo 2 compared to the halo 1. The sound speed ($\rm c_s$) is about 10 km/s and follows the temperature profile. To assess the scale height of a rotationally supported structure which is expected to form as a consequence of angular momentum conservation \citep{Oh2002,Volonteri2005,LatifVolonteri15}, we computed  V$_{rot}/ \sigma$ where V$_{rot}$ is the rotational velocity and $\sigma$ is the velocity dispersion ($\sqrt{c_s^2 + V_{turb}^2}$). V$_{rot}/ \sigma$  is inversely proportional to the scale height (H/R) and is shown in figure \ref{fig}. The ratio of V$_{rot}/ \sigma$ is about $\sim$ 0.7  for halo 1 and 1.0 for the halo 2. In the very central  part of the halo V$_{rot}/ \sigma$ is higher for halo 2 while it declines for halo 1. These estimates of  V$_{rot}/ \sigma$  suggest that a thick disk-like structure is formed in each halo and is partly supported by thermal pressure.

Overall, both halos in our study show a similar behaviour for the chemical and physical properties.  However, the differences in the dynamical properties are  a consequence of higher turbulent energy (i.e. higher kinetic energy) in halo 1 and enhanced rotation at small scales in halo 2. We discuss their implications in the section below. In general, our estimates of rotational, turbulent and radial infall velocities are comparable with previous studies exploring the collapse under similar conditions \citep{Latif2013c,Regan2014a,Bcerra2014}. Particularly, the mass inflow rate is $\rm 0.1-1.0~M_{\odot}/yr$ as found in the above mentioned studies and meets the requirement for the formation of a supermassive star \citep{Schleicher13,Hosokawa2013}.

% Our results in this regime can not be fully trusted as we approach the resolution limits.
\subsection{Fragmentation}
\begin{figure*}
\includegraphics[scale=0.3]{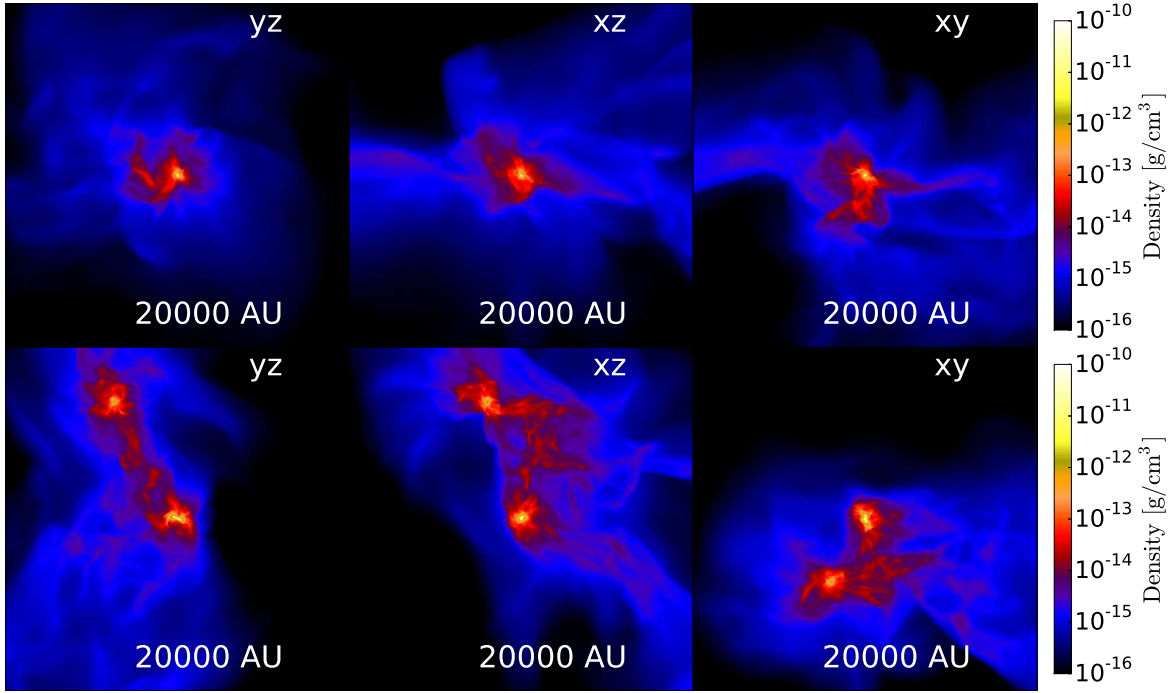}
\caption{The average density along the projection axis at the end stage of our simulations. The rows represent  halo 1 and  halo 2 while the columns from left to right show projections along x, y and z axis for the central 0.1 pc. The figure shows that only single object forms in the halo 1 while halo 2 shows two clumps.  The masses of the clumps are  a few $\rm M_{\odot}$ at this stage but are expected to grow rapidly at later times.}
\label{fig3}
\end{figure*}

\begin{figure*}
\includegraphics[scale=0.3]{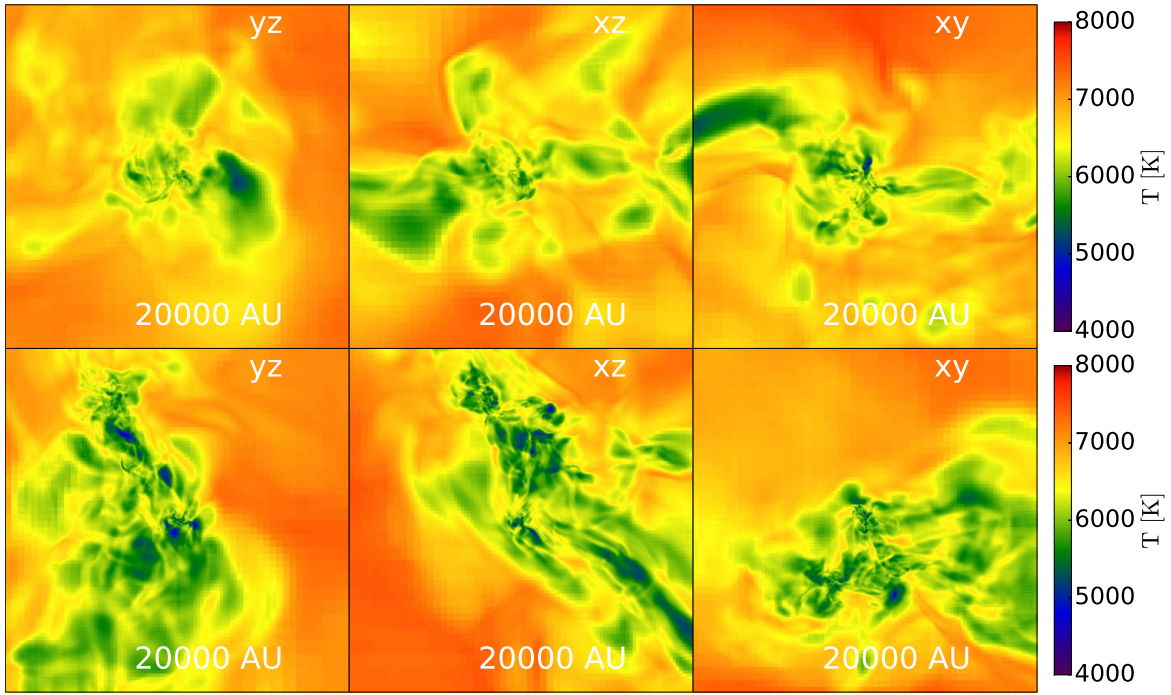}  
\caption{ Same as figure \ref{fig3}, showing the average temperature along the projection axis. The figure shows the local impact of $\rm H^-$ cooling.}
\label{fig4}
\end{figure*}

\begin{figure*}
\includegraphics[scale=0.3]{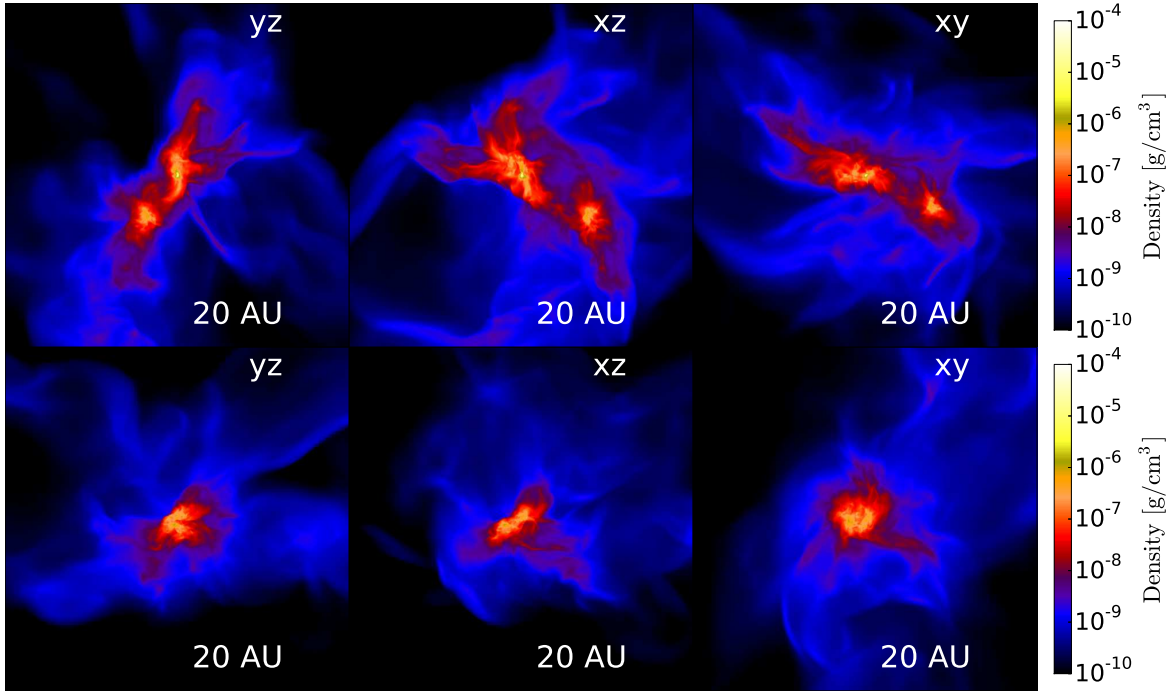} 
\caption{The average density along the projection axis at the end stage of our simulations. The rows represent  halo 1 and  halo 2 while the columns from left to right show projections along x, y and z axis for the central 20 AU part of both halos. It shows that on very small scales the halo 1 is fragmented into two clumps while halo 2 has only single clump at its centre.}
\label{fig5}
\end{figure*}

We found that  no fragmentation occurs above  a parsec scale in both halos which is in agreement with previous simulations \citep{Latif2013c,Regan2014a,Bcerra2014}.  While  within the central 0.1 pc, as shown in figure \ref{fig3}, halo 2 fragments into two clumps while halo 1 has only a single clump on this scale.  The clump masses in halo 2 are about 4 $\rm M_{\odot}$ and 2 $\rm M_{\odot}$ while the clump in halo 1 has  a mass of about 5 $\rm M_{\odot}$. Fragmentation at these scales is due to $\rm H^-$ cooling which becomes effective in this regime and  induces fragmentation in halo 2. It is further evident from figure \ref{fig0} that  the $\rm H^-$ cooling is more prominent in halo 2.  Similarly, the decrease in the temperature is visible in the corresponding temperature map shown in figure \ref{fig4}. The clumps in halo 2 are  about 8000 AU apart and may evolve into a binary at later stages.  The density structure at even smaller scales, i.e. the central 20 AU, is shown in figure \ref{fig5}. The cloud structure in  halo 1 is more filamentary at these scales and two fragments of 0.9 $\rm M_{\odot}$ and 0.4 $\rm M_{\odot}$ are formed within the central few AU.  In the case of halo 2, no fragmentation is observed on these scales and a disk-like structure is formed at its centre. There is more rotation in halo 2 at these scales as shown in V$_{rot}/ \sigma $ plot which leads to the formation of a disk. In the halo 1 there is less rotation and no disk like structure is formed on the scales of  a few AU.%although they switched off cooling to mimic the formation of a protostar. They were able to evolve simulation for  15 years after the formation of central protostar and found that disk around central star fragments into multiple clumps. While due to the higher spatial (as well as Jeans resolution) resolution and detailed chemical network  we could not do so. We stopped our simulation at a peak density of  $\rm 2 \times 10^{-3}~g/cm^3$ and $\rm \sim 10^{-5}~g/cm^3$ in halo 1 and halo 2, respectively.  We discuss below the possibility of additional fragmentation at later times if it occurs and its consequences.
 
\begin{figure*}
\includegraphics[scale=0.7]{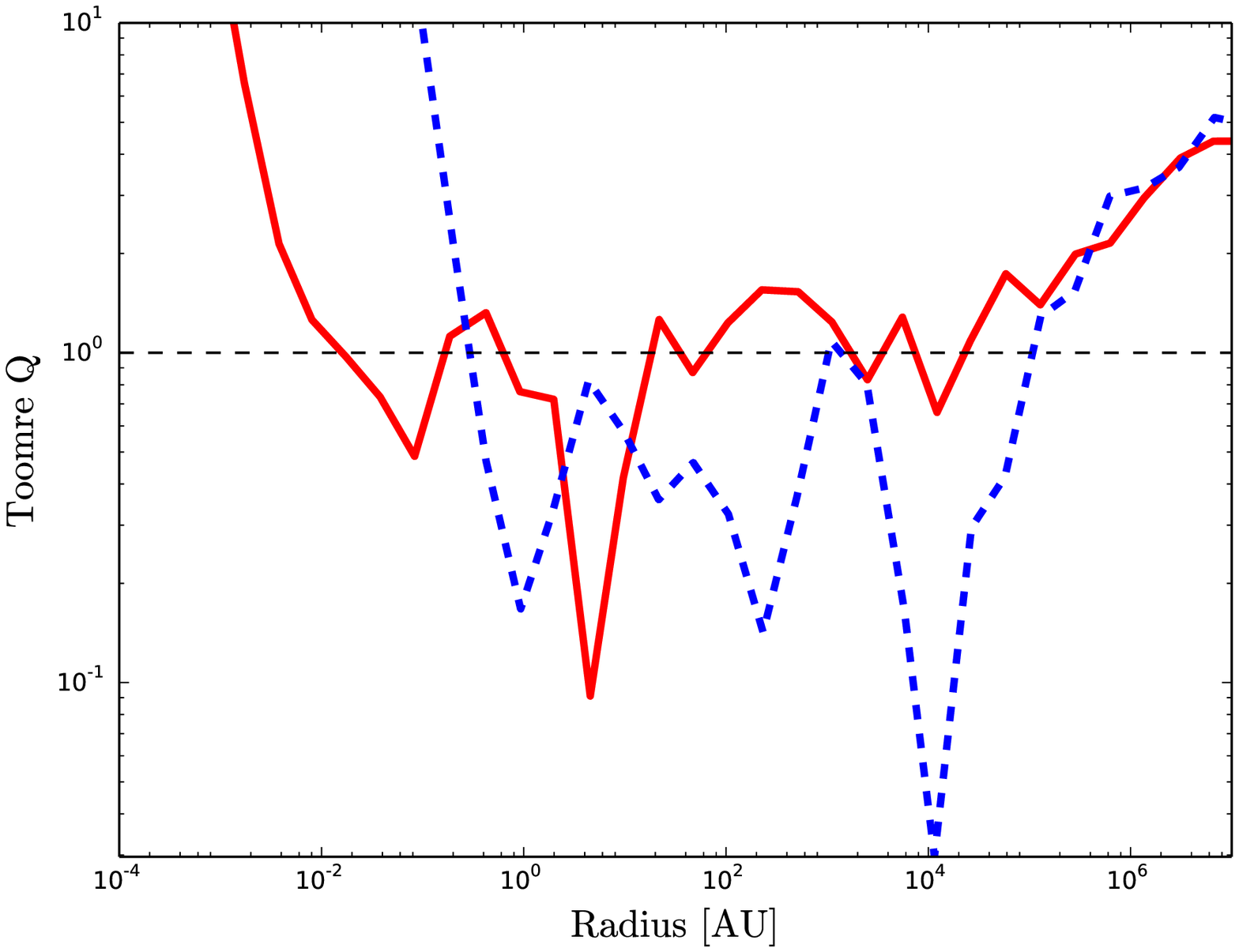} 
\caption{Toomre instability  parameter ''Q" for both halos at the end stage of our simulations. The red and blue lines represent halo1 and halo 2, respectively. The plot shows that  disks in the  halos become marginally unstable  between 1-10,000 AU due to the H$^-$ induced cooling. However, overall no strong fragmentation is expected.}
\label{fig2}
\end{figure*}

To  assess the stability of the disk as suggested by the estimates of V$_{rot}/ \sigma$ in both halos, we computed the Toomre stability parameter Q \citep{Toomre1964}.  The Toomre Q is calculated as  Q $\simeq \frac{ \sigma \Omega}{\pi G \Sigma}$  where $\sigma$ is the velocity dispersion, $\Omega$ is the rotational frequency, G is the gravitational constant and $\Sigma$ is the surface density of the gas. Our estimates of  Toomre Q are shown in figure \ref{fig2}. The disk is stable if Q$\geq$ 1 and unstable otherwise.  We found that the disk is  stable below 1 AU and above $\rm 10^5$ AU while in between  it is  marginally stable for both halos. The  wiggles in the plot are due to the substructure present inside the  disk. Particularly, the sharp dip in the value of Toomre Q at $\rm10^4$  AU corresponds to the additional clump in the halo 2  and similarly in the halo 1 due to the clump  at 1 AU scale. The stability of the disk on very small scales arises from the trapping of cooling radiation and the marginally stable regime is a consequence of $\rm H^-$ cooling. Indeed, $\rm H^-$ cooling in this regime lead to fragmentation in halo 2 where clumps are separated by a few thousand AU. We expect no further fragmentation on smaller scales as all cooling channels are shut off and the collapse is expected to proceed adiabatically. However, fragmentation between 1 AU and $\rm 10^4$ AU may still occur at later times. Nevertheless, the clumps forming in the central 100 AU may migrate inward and merge with the central clump \citep{Inayoshi2014b,Latif2015Disk,LatifViscous2015}. On the other hand, fragmentation at larger scales as seen in halo 2 may lead to the formation of a binary system. 

 To summarise, our results suggest that $\rm H^-$ cooling can induce fragmentation at a few 1000 AU such as in halo 2  but does not prevent the formation of a supermassive star at the center of atomic cooling halos where molecular hydrogen cooling remains suppressed. However, the impact of $\rm H^-$ cooling may vary from halo to halo. Halo 1 has a higher turbulent energy which prevents fragmentation at larger scales (around a few thousand AU).  The higher rotation at small scales (a few AU) in  halo 2 helps in the formation of a stable disk at small scales, the disk around the central star in halo 1 is expected to form during the later stages of collapse. We note that fragmentation in the previous studies without  $\rm H^-$ cooling \citep{Latif2013c,Regan2014a,Bcerra2014} occurred due to the unstable disk around the central protostar on scales of the order of $\sim$ 100 AU  at later times. Due to their simplified chemical networks and the lower spatial resolution these studies were able to evolve the simulations for a few dynamical timescales and therefore the central clumps masses were higher compared to the present study.  The masses of the central clumps in both halos are  a few solar masses but are expected to grow rapidly due to the presence of large accretion flows. We anticipate that the central clumps may experience intermittent accretion due to the clumpy medium inside the halo and the migration of additional clumps at later times.  We further expect  that the central clump may reach $\rm \sim 10^5~M_{\odot}$ within a Myr. Overall, fragmentation induced by $\rm H^-$ cooling  is different from the disk fragmentation around a protostar as the former occurs at scales where the clump migration time is considerably longer, potentially leading to the formation of a binary.

\section{ Discussion and conclusions}
We performed high resolution cosmological simulations for two halos of a few times $\rm 10^7~M_{\odot}$  at z >10 by employing a  detailed chemical model which includes  all the relevant chemical processes mentioned in the section 2. We  presumed that the halos are metal free and illuminated by a strong LW flux emitted from a nearby star forming galaxy. In this study, we particularly explored the impact of $\rm H^-$ cooling and employed realistic opacities both for the $\rm H^-$ and the collisional excitation cooling.  By exploiting the adaptive mesh refinement technique, we resolved the collapse down to the scales of $\rm 10^{-4}$ AU and up to unprecedented densities of $\rm \sim 10^{-3}~g/cm^{3}$.

Our results show that  collapse proceeds isothermally with a temperature of 8000 K up to densities of $\rm 10^{-16}~g/cm^{3}$, $\rm H^-$ cooling becomes effective between densities of $\rm 10^{-16}-10^{-8}~g/cm^{3}$, and brings the gas temperature  down to 5000 K.  Above densities of $\rm 10^{-6} ~g/cm^{3}$, the gas starts to become optical thick both to the bound-free $\rm H^-$ emission and the Rayleigh scattering of H atoms. Consequently, the temperature rises but the collisional ionisation cooling operates up to densities of about $\rm 10^{-4}~g/cm^{3}$ and and keeps the temperature  at about $\rm 10^4$ K.  At even higher densities, the collapse is expected to become completely adiabatic and a supermassive protostar is expected to form at this stage. The gas remains largely neutral during the collapse,  the degree of ionisation slightly decreases during the $\rm H^-$ cooling phase and starts to increase again as the gas becomes optically thick to cooling radiation.

The density  follows an $\rm R^{-2}$ profile as expected from an isothermal collapse.  Large inflow rates of about $\rm 1~M_{\odot}/yr$ are observed down to  a scale of $\rm 10^{-3}$ AU. Our estimates of V$_{rot}/ \sigma$ show that a rotationally supported disk-like structure with scale height of  0.6 is formed in both halos.  $\rm H^-$ cooling induces fragmentation in halo 2  at scales of about 8000 AU which  later may lead to binary formation. Moreover, fragmentation also occurs even on  a scale of about 1 AU in halo 1. Overall, our results suggest that although fragmentation occasionally occurs, $\rm H^-$ cooling does not prevent the formation of massive objects. The  trapping of cooling radiation enhances the gas temperature below 1 AU and  stabilises the collapse. No fragmentation is expected below the resolved scales but  fragmentation may occur above 1 AU at later times but we expect the clumps to get merged on shorter time scales \citep{Latif2015Disk,LatifViscous2015}.

In one of the two simulated halos $\rm H^- $ cooling induces fragmentation  at scales of about 8000 AU which may later form a binary system.  Our simulations do not allow us to estimate how often a binary formation may take place as our sample size is too small and due to the exorbitant computational costs simulations could not be evolved for longer time scale. In the future, such simulations should be performed for a larger sample of halos to estimate how often such conditions can occur. Our calculations approximately take into account the bound-free $\rm H^-$ absorption and the Rayleigh scattering  opacities.  Three dimensional cosmological hydrodynamics  simulations employing complete radiative transfer  should be performed to better understand the role  of radiation trapping during the direct collapse.

The comparison of our study with simulations employing adhoc initial conditions suggests that  the chemical and thermal properties in both cases are similar due to the self-similar behaviour of  the run-away collapse. However, dynamical properties such as  the turbulent velocity dispersion, angular momentum and rotational velocities are different. The complex interplay between turbulence, rotation, thermodynamics and  angular momentum transfer can lead to  differences in  fragmentation. The role of the initial conditions becomes more relevant after the formation of an  accretion disk and requires realistic cosmological simulations to take into account these effects. Moreover, simulations with simplified non-cosmological initial conditions may miss gravitational torques from the neighbouring halos and large scale inflows which are important for the long term evolution of the central objects in these halos.

We employed here a strong LW flux of strength $\rm J_{21}=10^5$ which is required to ensure  an isothermal collapse in  the halos  studied here.  Such strength of the LW flux in combination with metal free halos makes the sites of DCBHs quite rare \citep{Dijksta2014,Inayoshi2015,Habouzit2015}. However,  recent studies by \cite{LatifVolonteri15},  \cite{LatifViscous2015} and \cite{Schleicher2015} suggest that a completely isothermal collapse may not always be necessary to form massive black holes seeds.

%\begin{figure*}
%\includegraphics[scale=0.3]{Temperature_Projection.ps}  
%\caption{ Same as figure \ref{fig5}, showing the average temperature along the projection axis. The figure shows that the densest clumps of gas have a temperature of $\rm \sim 10^4$ K due to the trapping of the cooling radiation.}
%\label{fig6}
%\end{figure*}

 \section*{Acknowledgments}
We thank Marta Volonteri for interesting discussions. This project has received funding from the European Union's Horizon 2020 research and innovation programme under the Marie Sklodowska-Curie grant  agreement N$^o$ 656428. This work was granted access to the HPC resources of TGCC under the allocation x2015046955 made by GENCI. The research leading to these results has also received funding from the European Research Council under the European Community's Seventh Framework Programme (FP7/2007-2013 Grant Agreement no. 614199, project ``BLACK''). The simulation results are analyzed using the visualization toolkit for astrophysical data YT  \citep{Turk2011}. We also thank the anonymous referee for helpful suggestions.

%The research leading to these results has received funding from the European Research Council under the European Community's Seventh Framework Programme (FP7/2007-2013 Grant Agreement no. 614199, project ``BLACK'').
\bibliography{smbhs.bib}

\begin{thebibliography}{}

\bibitem[\protect\citeauthoryear{{Agarwal} \& {Khochfar}}{{Agarwal} \&
  {Khochfar}}{2015}]{Agarwal2015}
{Agarwal} B.,  {Khochfar} S.,  2015, \mnras, 446, 160

\bibitem[\protect\citeauthoryear{{Alvarez}, {Wise} \& {Abel}}{{Alvarez}
  et~al.}{2009}]{Alvarez2009}
{Alvarez} M.~A.,  {Wise} J.~H.,    {Abel} T.,  2009, \apjl, 701, L133

\bibitem[\protect\citeauthoryear{{Ba{\~n}ados} et~al.,}{{Ba{\~n}ados}
  et~al.}{2014}]{Banados2014}
{Ba{\~n}ados} E.  et~al., 2014, \aj, 148, 14

\bibitem[\protect\citeauthoryear{{Baumgarte} \& {Shapiro}}{{Baumgarte} \&
  {Shapiro}}{1999}]{Baumgarte1999}
{Baumgarte} T.~W.,  {Shapiro} S.~L.,  1999, \apj, 526, 941

\bibitem[\protect\citeauthoryear{{Becerra}, {Greif}, {Springel} \&
  {Hernquist}}{{Becerra} et~al.}{2015}]{Bcerra2014}
{Becerra} F.,  {Greif} T.~H.,  {Springel} V.,    {Hernquist} L.~E.,  2015,
  \mnras, 446, 2380

\bibitem[\protect\citeauthoryear{{Begelman}}{{Begelman}}{2010}]{Begelman2010}
{Begelman} M.~C.,  2010, \mnras, 402, 673

\bibitem[\protect\citeauthoryear{{Begelman} \& {Shlosman}}{{Begelman} \&
  {Shlosman}}{2009}]{Begelman2009}
{Begelman} M.~C.,  {Shlosman} I.,  2009, \apjl, 702, L5

\bibitem[\protect\citeauthoryear{{Begelman}, {Volonteri} \& {Rees}}{{Begelman}
  et~al.}{2006}]{Begelman2006}
{Begelman} M.~C.,  {Volonteri} M.,    {Rees} M.~J.,  2006, \mnras, 370, 289

\bibitem[\protect\citeauthoryear{{Bromm} \& {Loeb}}{{Bromm} \&
  {Loeb}}{2003}]{Brom2003}
{Bromm} V.,  {Loeb} A.,  2003, \apj, 596, 34

\bibitem[\protect\citeauthoryear{{Bryan} et~al.,}{{Bryan}
  et~al.}{2014}]{Enzocode2014}
{Bryan} G.~L.  et~al., 2014, \apjs, 211, 19

\bibitem[\protect\citeauthoryear{{Choi}, {Shlosman} \& {Begelman}}{{Choi}
  et~al.}{2015}]{Choi2015}
{Choi} J.-H.,  {Shlosman} I.,    {Begelman} M.~C.,  2015, \mnras, 450, 4411

\bibitem[\protect\citeauthoryear{{Devecchi} \& {Volonteri}}{{Devecchi} \&
  {Volonteri}}{2009}]{Devecchi2009}
{Devecchi} B.,  {Volonteri} M.,  2009, \apj, 694, 302

\bibitem[\protect\citeauthoryear{{Dijkstra}, {Ferrara} \&
  {Mesinger}}{{Dijkstra} et~al.}{2014}]{Dijksta2014}
{Dijkstra} M.,  {Ferrara} A.,    {Mesinger} A.,  2014, \mnras, 442, 2036

\bibitem[\protect\citeauthoryear{{Fan}, {Strauss}, {Richards}, {Hennawi},
  {Becker}, {White} \& {Diamond-Stanic}}{{Fan} et~al.}{2006}]{Fan2006}
{Fan} X.,  {Strauss} M.~A.,  {Richards} G.~T.,  {Hennawi} J.~F.,  {Becker}
  R.~H.,  {White} R.~L.,    {Diamond-Stanic} A.~M.,  2006, \aj, 131, 1203

\bibitem[\protect\citeauthoryear{{Ferrara}, {Salvadori}, {Yue} \&
  {Schleicher}}{{Ferrara} et~al.}{2014}]{Ferrara14}
{Ferrara} A.,  {Salvadori} S.,  {Yue} B.,    {Schleicher} D.,  2014, \mnras,
  443, 2410

\bibitem[\protect\citeauthoryear{{Glover}}{{Glover}}{2015}]{Glover2015b}
{Glover} S.~C.~O.,  2015, ArXiv e-prints:1504.00514

\bibitem[\protect\citeauthoryear{{Grassi}, {Bovino}, {Schleicher}, {Prieto},
  {Seifried}, {Simoncini} \& {Gianturco}}{{Grassi} et~al.}{2014}]{Grassi2014}
{Grassi} T.,  {Bovino} S.,  {Schleicher} D.~R.~G.,  {Prieto} J.,  {Seifried}
  D.,  {Simoncini} E.,    {Gianturco} F.~A.,  2014, \mnras, 439, 2386

\bibitem[\protect\citeauthoryear{{Habouzit} et~al.,}{{Habouzit}
  et~al.}{2015}]{Habouzit2015}
{Habouzit} M.  et~al., 2015, ArXiv e-prints:1507.05971

\bibitem[\protect\citeauthoryear{{Haiman}}{{Haiman}}{2013}]{Haiman2013}
{Haiman} Z.,  2013, in {Wiklind} T.,  {Mobasher} B.,   {Bromm} V.,  eds,
  Astrophysics and Space Science Library Vol. 396, Astrophysics and Space
  Science Library. p.~293

\bibitem[\protect\citeauthoryear{{Hartwig}, {Glover}, {Klessen}, {Latif} \&
  {Volonteri}}{{Hartwig} et~al.}{2015}]{Hartwig2015}
{Hartwig} T.,  {Glover} S.~C.~O.,  {Klessen} R.~S.,  {Latif} M.~A.,
  {Volonteri} M.,  2015, \mnras, 452, 1233

\bibitem[\protect\citeauthoryear{{Hosokawa}, {Yorke}, {Inayoshi}, {Omukai} \&
  {Yoshida}}{{Hosokawa} et~al.}{2013}]{Hosokawa2013}
{Hosokawa} T.,  {Yorke} H.~W.,  {Inayoshi} K.,  {Omukai} K.,    {Yoshida} N.,
  2013, \apj, 778, 178

\bibitem[\protect\citeauthoryear{{Inayoshi} \& {Haiman}}{{Inayoshi} \&
  {Haiman}}{2014}]{Inayoshi2014b}
{Inayoshi} K.,  {Haiman} Z.,  2014, \mnras, 445, 1549

\bibitem[\protect\citeauthoryear{{Inayoshi}, {Omukai} \& {Tasker}}{{Inayoshi}
  et~al.}{2014}]{Inayoshi2014}
{Inayoshi} K.,  {Omukai} K.,    {Tasker} E.,  2014, \mnras, 445, L109

\bibitem[\protect\citeauthoryear{{Inayoshi} \& {Tanaka}}{{Inayoshi} \&
  {Tanaka}}{2015}]{Inayoshi2015}
{Inayoshi} K.,  {Tanaka} T.~L.,  2015, \mnras, 450, 4350

\bibitem[\protect\citeauthoryear{{Jarosik}, {Bennett}, {Dunkley}, {Gold},
  {Greason}, {Halpern}, {Hill} \& {Hinshaw}}{{Jarosik}
  et~al.}{2011}]{Jarosik2011}
{Jarosik} N.,  {Bennett} C.~L.,  {Dunkley} J.,  {Gold} B.,  {Greason} M.~R.,
  {Halpern} M.,  {Hill} R.~S.,    {Hinshaw} G.,  2011, \apjs, 192, 14

\bibitem[\protect\citeauthoryear{{Johnson} \& {Bromm}}{{Johnson} \&
  {Bromm}}{2007}]{Johnson2007}
{Johnson} J.~L.,  {Bromm} V.,  2007, \mnras, 374, 1557

\bibitem[\protect\citeauthoryear{{Johnson}, {Whalen}, {Li} \& {Holz}}{{Johnson}
  et~al.}{2013}]{Johnson2013b}
{Johnson} J.~L.,  {Whalen} D.~J.,  {Li} H.,    {Holz} D.~E.,  2013, \apj, 771,
  116

\bibitem[\protect\citeauthoryear{{Larson}}{{Larson}}{1969}]{Larson1969}
{Larson} R.~B.,  1969, \mnras, 145, 271

\bibitem[\protect\citeauthoryear{{Latif}, {Bovino}, {Grassi}, {Schleicher} \&
  {Spaans}}{{Latif} et~al.}{2015}]{Latif2015a}
{Latif} M.~A.,  {Bovino} S.,  {Grassi} T.,  {Schleicher} D.~R.~G.,    {Spaans}
  M.,  2015, \mnras, 446, 3163

\bibitem[\protect\citeauthoryear{{Latif}, {Bovino}, {Van Borm}, {Grassi},
  {Schleicher} \& {Spaans}}{{Latif} et~al.}{2014}]{Latif2014UV}
{Latif} M.~A.,  {Bovino} S.,  {Van Borm} C.,  {Grassi} T.,  {Schleicher}
  D.~R.~G.,    {Spaans} M.,  2014, \mnras, 443, 1979

\bibitem[\protect\citeauthoryear{{Latif}, {Omukai}, {Habouzit}, {Schleicher} \&
  {Volonteri}}{{Latif} et~al.}{2015}]{Latif2015dust}
{Latif} M.~A.,  {Omukai} K.,  {Habouzit} M.,  {Schleicher} D.~R.~G.,
  {Volonteri} M.,  2015, ArXiv e-prints:1509.07034

\bibitem[\protect\citeauthoryear{{Latif} \& {Schleicher}}{{Latif} \&
  {Schleicher}}{2015a}]{Latif2015Disk}
{Latif} M.~A.,  {Schleicher} D.~R.~G.,  2015a, \mnras, 449, 77

\bibitem[\protect\citeauthoryear{{Latif} \& {Schleicher}}{{Latif} \&
  {Schleicher}}{2015b}]{LatifViscous2015}
{Latif} M.~A.,  {Schleicher} D.~R.~G.,  2015b, \aap, 578, A118

\bibitem[\protect\citeauthoryear{{Latif}, {Schleicher} \& {Schmidt}}{{Latif}
  et~al.}{2014}]{LatifMag2014}
{Latif} M.~A.,  {Schleicher} D.~R.~G.,    {Schmidt} W.,  2014, \mnras, 440,
  1551

\bibitem[\protect\citeauthoryear{{Latif}, {Schleicher}, {Schmidt} \&
  {Niemeyer}}{{Latif} et~al.}{2013a}]{Latif2013c}
{Latif} M.~A.,  {Schleicher} D.~R.~G.,  {Schmidt} W.,    {Niemeyer} J.,  2013a,
  \mnras, 433, 1607

\bibitem[\protect\citeauthoryear{{Latif}, {Schleicher}, {Schmidt} \&
  {Niemeyer}}{{Latif} et~al.}{2013b}]{LatifPopIII13}
{Latif} M.~A.,  {Schleicher} D.~R.~G.,  {Schmidt} W.,    {Niemeyer} J.,  2013b,
  \apjl, 772, L3

\bibitem[\protect\citeauthoryear{{Latif}, {Schleicher}, {Schmidt} \&
  {Niemeyer}}{{Latif} et~al.}{2013c}]{Latifdynamo2013}
{Latif} M.~A.,  {Schleicher} D.~R.~G.,  {Schmidt} W.,    {Niemeyer} J.,  2013c,
  \mnras, 432, 668

\bibitem[\protect\citeauthoryear{{Latif}, {Schleicher}, {Schmidt} \&
  {Niemeyer}}{{Latif} et~al.}{2013d}]{Latif2013d}
{Latif} M.~A.,  {Schleicher} D.~R.~G.,  {Schmidt} W.,    {Niemeyer} J.~C.,
  2013d, \mnras, 436, 2989

\bibitem[\protect\citeauthoryear{{Latif} \& {Volonteri}}{{Latif} \&
  {Volonteri}}{2015}]{LatifVolonteri15}
{Latif} M.~A.,  {Volonteri} M.,  2015, \mnras, 452, 1026

\bibitem[\protect\citeauthoryear{{Latif}, {Zaroubi} \& {Spaans}}{{Latif}
  et~al.}{2011}]{Latif2011}
{Latif} M.~A.,  {Zaroubi} S.,    {Spaans} M.,  2011, \mnras, 411, 1659

\bibitem[\protect\citeauthoryear{{Lupi}, {Colpi}, {Devecchi}, {Galanti} \&
  {Volonteri}}{{Lupi} et~al.}{2014}]{Lupi2014}
{Lupi} A.,  {Colpi} M.,  {Devecchi} B.,  {Galanti} G.,    {Volonteri} M.,
  2014, \mnras, 442, 3616

\bibitem[\protect\citeauthoryear{{Madau}, {Haardt} \& {Dotti}}{{Madau}
  et~al.}{2014}]{Madau2014}
{Madau} P.,  {Haardt} F.,    {Dotti} M.,  2014, \apjl, 784, L38

\bibitem[\protect\citeauthoryear{{Mortlock} et~al.,}{{Mortlock}
  et~al.}{2011}]{MOrtlock2011}
{Mortlock} D.~J.  et~al., 2011, \nat, 474, 616

\bibitem[\protect\citeauthoryear{{Oh} \& {Haiman}}{{Oh} \&
  {Haiman}}{2002}]{Oh2002}
{Oh} S.~P.,  {Haiman} Z.,  2002, \apj, 569, 558

\bibitem[\protect\citeauthoryear{{Omukai}}{{Omukai}}{2000}]{Omukai2000}
{Omukai} K.,  2000, ApJ, 534, 809

\bibitem[\protect\citeauthoryear{{Omukai}}{{Omukai}}{2001}]{Omukai2001}
{Omukai} K.,  2001, \apj, 546, 635

\bibitem[\protect\citeauthoryear{{O'Shea}, {Bryan}, {Bordner}, {Norman},
  {Abel}, {Harkness} \& {Kritsuk}}{{O'Shea} et~al.}{2004}]{OShea2004}
{O'Shea} B.~W.,  {Bryan} G.,  {Bordner} J.,  {Norman} M.~L.,  {Abel} T.,
  {Harkness} R.,    {Kritsuk} A.,  2004, ArXiv Astrophysics e-prints 0403044

\bibitem[\protect\citeauthoryear{{Pacucci}, {Volonteri} \& {Ferrara}}{{Pacucci}
  et~al.}{2015}]{Pacucci2015}
{Pacucci} F.,  {Volonteri} M.,    {Ferrara} A.,  2015, \mnras, 452, 1922

\bibitem[\protect\citeauthoryear{{Penston}}{{Penston}}{1969}]{Penston1969}
{Penston} M.~V.,  1969, \mnras, 144, 425

\bibitem[\protect\citeauthoryear{{Prieto}, {Jimenez} \& {Haiman}}{{Prieto}
  et~al.}{2013}]{Prieto2013}
{Prieto} J.,  {Jimenez} R.,    {Haiman} Z.,  2013, \mnras, 436, 2301

\bibitem[\protect\citeauthoryear{{Rees}}{{Rees}}{1978}]{Rees78}
{Rees} M.~J.,  1978, The Observatory, 98, 210

\bibitem[\protect\citeauthoryear{{Regan} \& {Haehnelt}}{{Regan} \&
  {Haehnelt}}{2009}]{Regan09}
{Regan} J.~A.,  {Haehnelt} M.~G.,  2009, \mnras, 393, 858

\bibitem[\protect\citeauthoryear{{Regan}, {Johansson} \& {Haehnelt}}{{Regan}
  et~al.}{2014}]{Regan2014a}
{Regan} J.~A.,  {Johansson} P.~H.,    {Haehnelt} M.~G.,  2014, \mnras, 439,
  1160

\bibitem[\protect\citeauthoryear{{Regan}, {Johansson} \& {Wise}}{{Regan}
  et~al.}{2014}]{Regan2014B}
{Regan} J.~A.,  {Johansson} P.~H.,    {Wise} J.~H.,  2014, \apj, 795, 137

\bibitem[\protect\citeauthoryear{{Saigo} \& {Hanawa}}{{Saigo} \&
  {Hanawa}}{1998}]{Saigo1998}
{Saigo} K.,  {Hanawa} T.,  1998, \apj, 493, 342

\bibitem[\protect\citeauthoryear{{Sakurai}, {Hosokawa}, {Yoshida} \&
  {Yorke}}{{Sakurai} et~al.}{2015}]{Sakurai2015}
{Sakurai} Y.,  {Hosokawa} T.,  {Yoshida} N.,    {Yorke} H.~W.,  2015, \mnras,
  452, 755

\bibitem[\protect\citeauthoryear{{Schleicher}, {Bovino}, {Latif}, {Ferrara} \&
  {Grassi}}{{Schleicher} et~al.}{2015}]{Schleicher2015}
{Schleicher} D.~R.~G.,  {Bovino} S.,  {Latif} M.~A.,  {Ferrara} A.,    {Grassi}
  T.,  2015, ArXiv e-prints:1504.06296

\bibitem[\protect\citeauthoryear{{Schleicher}, {Palla}, {Ferrara}, {Galli} \&
  {Latif}}{{Schleicher} et~al.}{2013}]{Schleicher13}
{Schleicher} D.~R.~G.,  {Palla} F.,  {Ferrara} A.,  {Galli} D.,    {Latif} M.,
  2013, \aap, 558, A59

\bibitem[\protect\citeauthoryear{{Schleicher}, {Spaans} \&
  {Glover}}{{Schleicher} et~al.}{2010}]{Schleicher10}
{Schleicher} D.~R.~G.,  {Spaans} M.,    {Glover} S.~C.~O.,  2010, \apjl, 712,
  L69

\bibitem[\protect\citeauthoryear{{Shang}, {Bryan} \& {Haiman}}{{Shang}
  et~al.}{2010}]{Shang2010}
{Shang} C.,  {Bryan} G.~L.,    {Haiman} Z.,  2010, \mnras, 402, 1249

\bibitem[\protect\citeauthoryear{{Shlosman}, {Choi}, {Begelman} \&
  {Nagamine}}{{Shlosman} et~al.}{2015}]{Shlosman2015}
{Shlosman} I.,  {Choi} J.-H.,  {Begelman} M.~C.,    {Nagamine} K.,  2015, ArXiv
  e-prints

\bibitem[\protect\citeauthoryear{{Spaans} \& {Silk}}{{Spaans} \&
  {Silk}}{2006}]{Spaans2006}
{Spaans} M.,  {Silk} J.,  2006, \apj, 652, 902

\bibitem[\protect\citeauthoryear{{Sugimura}, {Omukai} \& {Inoue}}{{Sugimura}
  et~al.}{2014}]{Sugimura14}
{Sugimura} K.,  {Omukai} K.,    {Inoue} A.~K.,  2014, \mnras, 445, 544

\bibitem[\protect\citeauthoryear{{Toomre}}{{Toomre}}{1964}]{Toomre1964}
{Toomre} A.,  1964, \apj, 139, 1217

\bibitem[\protect\citeauthoryear{{Treister}, {Schawinski}, {Volonteri} \&
  {Natarajan}}{{Treister} et~al.}{2013}]{Treister2013}
{Treister} E.,  {Schawinski} K.,  {Volonteri} M.,    {Natarajan} P.,  2013,
  \apj, 778, 130

\bibitem[\protect\citeauthoryear{{Turk}, {Smith}, {Oishi}, {Skory}, {Skillman},
  {Abel} \& {Norman}}{{Turk} et~al.}{2011}]{Turk2011}
{Turk} M.~J.,  {Smith} B.~D.,  {Oishi} J.~S.,  {Skory} S.,  {Skillman} S.~W.,
  {Abel} T.,    {Norman} M.~L.,  2011, \apjs, 192, 9

\bibitem[\protect\citeauthoryear{{Van Borm}, {Bovino}, {Latif}, {Schleicher},
  {Spaans} \& {Grassi}}{{Van Borm} et~al.}{2014}]{VanBorm2014}
{Van Borm} C.,  {Bovino} S.,  {Latif} M.~A.,  {Schleicher} D.~R.~G.,  {Spaans}
  M.,    {Grassi} T.,  2014, \aap, 572, A22

\bibitem[\protect\citeauthoryear{{Venemans} et~al.,}{{Venemans}
  et~al.}{2013}]{Venemans2013}
{Venemans} B.~P.  et~al., 2013, \apj, 779, 24

\bibitem[\protect\citeauthoryear{{Venemans} et~al.,}{{Venemans}
  et~al.}{2015}]{Venemans2015}
{Venemans} B.~P.  et~al., 2015, \mnras, 453, 2259

\bibitem[\protect\citeauthoryear{{Visbal}, {Haiman} \& {Bryan}}{{Visbal}
  et~al.}{2014}]{Visbal2014}
{Visbal} E.,  {Haiman} Z.,    {Bryan} G.~L.,  2014, \mnras, 445, 1056

\bibitem[\protect\citeauthoryear{{Volonteri}}{{Volonteri}}{2010}]{Volonteri2010}
{Volonteri} M.,  2010, \aapr, 18, 279

\bibitem[\protect\citeauthoryear{{Volonteri}, {Lodato} \&
  {Natarajan}}{{Volonteri} et~al.}{2008}]{Volonteri2008}
{Volonteri} M.,  {Lodato} G.,    {Natarajan} P.,  2008, \mnras, 383, 1079

\bibitem[\protect\citeauthoryear{{Volonteri} \& {Rees}}{{Volonteri} \&
  {Rees}}{2005}]{Volonteri2005}
{Volonteri} M.,  {Rees} M.~J.,  2005, \apj, 633, 624

\bibitem[\protect\citeauthoryear{{Willott} et~al.,}{{Willott}
  et~al.}{2007}]{Willott2007}
{Willott} C.~J.  et~al., 2007, \aj, 134, 2435

\bibitem[\protect\citeauthoryear{{Wise}, {Turk} \& {Abel}}{{Wise}
  et~al.}{2008}]{Wise2008}
{Wise} J.~H.,  {Turk} M.~J.,    {Abel} T.,  2008, \apj, 682, 745

\bibitem[\protect\citeauthoryear{{Wolcott-Green}, {Haiman} \&
  {Bryan}}{{Wolcott-Green} et~al.}{2011}]{WolocottGreen2011}
{Wolcott-Green} J.,  {Haiman} Z.,    {Bryan} G.~L.,  2011, \mnras, 418, 838

\bibitem[\protect\citeauthoryear{{Wu} et~al.,}{{Wu} et~al.}{2015}]{Wu2015}
{Wu} X.-B.  et~al., 2015, \nat, 518, 512

\bibitem[\protect\citeauthoryear{{Yue}, {Ferrara}, {Salvaterra}, {Xu} \&
  {Chen}}{{Yue} et~al.}{2014}]{Yue2014}
{Yue} B.,  {Ferrara} A.,  {Salvaterra} R.,  {Xu} Y.,    {Chen} X.,  2014,
  \mnras, 440, 1263

\end{thebibliography}
 
\newpage

\end{document}